\documentclass[preprint, 12pt]{elsarticle}
\usepackage[monochrome]{xcolor}
\usepackage{graphicx}
\usepackage{subcaption}
\usepackage{amssymb, amsmath}

\biboptions{sort&compress}

\journal{ }

\usepackage{hyperref}

\begin{document}

\begin{frontmatter}

\title{GTApprox: surrogate modeling for industrial design}

\author{
Mikhail Belyaev\textsuperscript{1}, 
Evgeny Burnaev\textsuperscript{1,2},
Ermek Kapushev\textsuperscript{1,2}, 
Maxim Panov\textsuperscript{1,2}, \\
Pavel Prikhodko\textsuperscript{1}, 
Dmitry Vetrov\textsuperscript{2},
 and  
Dmitry Yarotsky\textsuperscript{1,2}}

\address{\textsuperscript{1}Kharkevich Institute for Information Transmission Problems, Bolshoy Karetny per. 19, build.1, Moscow 127051, Russia \\
\textsuperscript{2}Datadvance llc,  Nauchny proezd 17, Moscow 117246, Russia}

\begin{abstract}
We describe GTApprox --- a new tool for medium-scale surrogate modeling in industrial design. Compared to existing software, GTApprox brings several innovations: a few novel approximation algorithms, several advanced methods of automated model selection, novel options in the form of hints. We demonstrate the efficiency of GTApprox on a large collection of test problems. In addition, we describe several applications of GTApprox to real engineering problems.  
\end{abstract}


\end{frontmatter}

\section{Introduction}
\label{section: introduction}
Approximation problems (also known as regression problems) arise quite often in industrial design, and solutions of such problems are conventionally referred to as surrogate models~\cite{Forrester2008}. The most common application of surrogate modeling in engineering is in connection to engineering optimization~\cite{Simpson2008}. Indeed, on the one hand, design optimization plays a central role in the industrial design process; on the other hand, a single optimization step typically requires the optimizer to create or refresh a model of the response function whose optimum is sought, to be able to come up with a reasonable next design candidate. The surrogate models used in optimization range from simple local linear regression employed in the basic gradient-based optimization~\cite{Nesterov2004} to complex global models employed in the so-called Surrogate-Based Optimization (SBO)~\cite{Forrester2009}. Aside from optimization, surrogate modeling is used in dimension reduction~\cite{Xia2002,Zhu2006}, sensitivity analysis~\cite{Oakley2004,Sudret2008,Burnaev2015s,Burnaev2016s}, and for visualization of response functions. 

Mathematically, the approximation problem can generally be described as follows. We assume that we are given a finite sample of pairs $(\mathbf{x}_n, \mathbf{y}_n)_{n=1}^N$ (the ``training data''), where $\mathbf{x}_n\in\mathbb{R}^{d_\mathrm{in}}, \mathbf{y}_n\in\mathbb{R}^{d_\mathrm{out}}$. These pairs represent sampled inputs and outputs of an unknown response function $\mathbf{y} = f(\mathbf{x})$. 
Our goal is to construct a function (a surrogate model) $\widehat{f}:\mathbb{R}^{d_\mathrm{in}}\to \mathbb{R}^{d_\mathrm{out}}$ which should be as close as possible to the true function $f$.

A great variety of surrogate modeling methods exist, with different assumptions on the underlying response functions, data sets, and model structure~\cite{Hastie2010}. Bundled implementations of diverse surrogate modeling methods can be found in many software tools, for example in the excellent open-source general purpose statistical project R~\cite{R2012} and machine-learning Python library scikit-learn~\cite{Sklearn2011}, as well as in several engineering-oriented frameworks~\cite{Dakota2013,Hofmann2013,ModeFrontier}. Theoretically, any of these tools offers an engineer the necessary means to construct and use surrogate models covering a wide range of approximation scenarios. In practice, however, existing tools are often not very convenient to an engineer, for two main reasons.

1. {\bf Excessively technical user interface and its inconsistency across different surrogate modeling techniques. } Predictive modeling tools containing a variety of different modeling algorithms often provide a common top-level interface for loading training data and constructing and applying surrogate models. However, the algorithms themselves usually remain isolated; in particular, they typically have widely different sets of user options and tunable parameters. This is not surprising, as there is a substantial conceptual difference in the logic of different modeling methods. For example, standard linear regression uses a small number of fixed basis functions and only linear operations; kriging uses a large number of basis functions specifically adjusted to the training data and involves nonlinear parameters; artificial neural networks may use a variable set of basis functions and some elements of user-controlled self-validation; etc.  

Such isolation of algorithms requires the user to learn their properties in order to pick the right algorithm and to correctly set its options, and engineers rarely have time for that. An experienced researcher would know, for example, that artificial neural networks can produce quite accurate approximations for high-dimensional data, but when applied in 1D, the plotted results would almost invariably look very unconvincing (compared to, say, splines); kriging is a popular choice for moderately sized training sets, but will likely exhaust the on-board RAM if the training set has more than a few thousand elements; accurate approximations by neural networks may take several days to train; etc. In existing tools such expert knowledge is usually scattered in documentation, and users quite often resort to trial-and-error when choosing the algorithm.

2. {\bf Lack of attention to special features of engineering problems. } The bias of the engineering domain is already seen in the very fact that regression problems in industrial design are much more common than classification problems (i.e., those where one predicts a discrete label rather than a continuous value $\mathbf{y}\in\mathbb{R}^{d_\mathrm{out}}$), whereas quite the opposite seems to hold in the more general context of all commercial and scientific applications of predictive modeling\footnote{Note, for example, that classification problems form the majority of the 200+ Kaggle data mining contests \cite{Kaggle} and the 300+ UCI machine learning repository data sets \cite{Lichman:2013}, well reflecting current trends in this area.}. Moreover, the response function $f(\mathbf{x})$ considered in an engineering problem usually represents some physical quantity and is expected to vary smoothly or at least continuously with $\mathbf{x}$. At the same time, widely popular decision-tree-based methods such as random forests \cite{Breiman2001} and gradient boosting \cite{Friedman2001} produce discontinuous piece-wise constant surrogate models, completely inappropriate for, say, gradient-based optimization. This example is rather obvious and the issue can be solved by simply ignoring decision-tree-based methods, but, based on our experience of  surrogate modeling at industrial enterprises \cite{Grihon2013, Grihon2014, Belyaev2014, Struzik2013, datadvance_use_cases}, we can identify several more subtle elements of this engineering bias that require significant changes in the software architecture, in particular:
\begin{description} 
\item[Data anisotropy.] Training data can be very anisotropic with respect to different groups of variables. For example, a common source of data are experiments performed under different settings of parameters with some sort of detectors that have fixed positions (e.g., air pressure measured on a wing under different settings of Mach and angle of attack), and the surrogate model needs to predict the outcome of the experiment for a new setting of parameters and at a new detector position. It can easily be that the detectors are abundant and regularly distributed, while the number of experiments is scarce and their parameters are high-dimensional and irregularly scattered in the parameter space. If we only needed a surrogate model with respect to one of these two groups of input variables, we could easily point out an appropriate standard method (say, splines for detector position and regularized linear regression for the experiment parameters), but how to combine them into a single model? Such anisotropic scenarios, with different expected dependency properties, seem to be quite typical in the engineering domain~\cite{Montgomery2006}.

\item[Model smoothness and availability of gradients.]  As mentioned above, surrogate models in engineering are (more often than in other domains) used for optimization and sensitivity analysis, and are usually expected to reasonably smoothly depend on the input variables. Moreover, there is some trade-off between model smoothness and accuracy, so it is helpful to be able to directly control the amount of smoothness in the model. If a gradient-based optimization is to be applied to the model, it is beneficial to have the exact analytic gradient of the model, thus avoiding its expensive and often inaccurate numerical approximation.

\item[Local accuracy estimates.] Surrogate-based optimization requires, in addition to the approximation of the response function, a model estimating local accuracy of this approximation \cite{Jones1998,Forrester2009}. This model of local accuracy is very rarely provided in existing software, and is usually restricted to the method known in engineering literature as kriging \cite{Krige1951}, which has been recently paid much attention in machine learning community under the name of Gaussian process regression \cite{Rasmussen2005}.

\item[Handling multidimensional output.] In the literature, main attention is focused on modeling functions with a single scalar output~\cite{Forrester2008}. However, in engineering practice the output is very often multidimensional, i.e. the problem in question requires modeling several physical quantities as functions of input parameters. Especially challenging are situations when the outputs are highly correlated. An example is the modeling of pressure distribution along the airfoil as a function of airfoil shape. In such cases one expects the output components of a surrogate model to be accordingly correlated with each other.
\end{description}

In this note we describe a surrogate modeling tool GTApprox (Generic Tool for Approximation) designed with the goal of overcoming the above shortcomings of existing software. First, the tool contains multiple novel ``meta-algorithms'' providing the user with accessible means of controlling the process of modeling in terms of easily understood options, in addition to conventional method-specific parameters. Second, the tool has additional modes and features addressing the specific engineering needs pointed out above. 

Some algorithmic novelties of the tool have already been described earlier \cite{Belyaev2015, Belyaev2015b, Belyaev2016, Burnaev2013, Burnaev2015, Burnaev2015b, Burnaev2016, Burnaev2016b}; in the present paper we describe the tool as a whole, in particular focusing on the overall decision process and performance comparison that have not been published before. The tool is a part of the MACROS library~\cite{Burnaev2013b}. It can be used as a standalone Python module or with a GUI within the pSeven platform~\cite{Davydov2015}. The trial version of the tool is available at \cite{datadvance_downloads}. A very detailed exposition of the tool's functionality can be found in its Sphinx-based documentation \cite{datadvance_docs}.

The remainder of the paper is organized as follows. In Sections \ref{sec:algo} and \ref{sec:algo_sel} we describe the tool's structure and main algorithms. In particular, in Section \ref{sec:algo} we review individual approximation algorithms of the tool (such as splines, RSM, etc.), with the emphasis on novel elements and special features. In Section \ref{sec:algo_sel} we describe how the tool automatically chooses the appropriate individual algorithm for a given problem.
Next, in section \ref{sec:tests} we report results of comparison of the tool with alternative state-of-the-art surrogate modeling methods on a collection of test problems. Finally, in section \ref{sec:apps} we describe a few industrial applications of the tool.  

\section{Approximation algorithms and special features}\label{sec:algo}

\subsection{Approximation algorithms}
GTApprox is aimed at solving a wide range of approximation problems. There is no universal approximation algorithm which can efficiently solve all types of problems, so GTApprox contains many individual algorithms, that we hereafter refer to as \emph{techniques}, each providing the best approximation quality in a particular domain. Some of these techniques are more or less standard, while others are new or at least contain features rarely found in other software. We will briefly overview main individual techniques, focusing on their novelties useful for engineering design. 
  
\paragraph*{Response Surface Models (RSM)} This is a generalized linear regression including several approaches to estimation of regression coefficients. RSM can be either linear or quadratic with respect to input variables. Also, RSM supports categorical input variables. There are a number of ways to estimate unknown coefficients of RSM, among which GTApprox implements ridge regression\cite{Tikhonov1943}, stepwise regression~\cite{Efroymson1960} and the elastic net~\cite{Zou2005}.

\paragraph*{Splines With Tension (SPLT)} This is one-dimensional spline-based technique intended to combine the robustness of linear splines with the smoothness of cubic splines. A non-linear algorithm~\cite{Renka1987} is used for an adaptive selection of the optimal weights on each interval between neighboring points of DoE (Design of Experiment, i.e. the set of input vectors of the training set).
  
\paragraph*{Gaussian Processes (GP) and Sparse Gaussian Process (SGP)} These are flexible nonlinear techniques based on modeling training data as a realization of an infinite-dimensional Gaussian distribution defined by a mean function and a covariance function \cite{Cressie1993,Rasmussen2005}. GP allows us to construct approximations that exactly agree with the provided training data. Also, this technique provides local accuracy estimates based on the a posteriori covariance of the considered Gaussian process. Thanks to this important property we can use GP in surrogate-based optimization~\cite{Jones1998} and adaptive design of experiments~\cite{Burnaev2015}. In GTApprox, parameters of GP are optimized by a novel optimization algorithm with adaptive regularization~\cite{Burnaev2016}, improving the generalization ability of the approximation (see also results on theoretical properties of parameters estimates \cite{Burnaev2013c,Burnaev2013d,Burnaev2013dd}). GP memory requirements scale quadratically with the size of the training set, so this technique is not applicable to very large training sets. SGP is a version of GP that lifts this limitation by using only a suitably selected subset of training data and approximating the corresponding covariance matrices~\cite{Burnaev2015b}.      
 
\paragraph*{High Dimensional Approximation (HDA) and High Dimensional Approximation combined with Gaussian Processes (HDAGP)} HDA is a nonlinear, adaptive technique using decomposition over linear and nonlinear base functions from a functional dictionary. This technique is related to artificial neural networks and, more specifically, to the two-layer perceptron with a nonlinear activation function~\cite{Haykin1998}. However, neural networks are notorious for overfitting~\cite{Caruana2001} and for the need to adjust their parameters by trial-and-error. HDA contains many tweaks and novel structural elements intended to automate training and reduce overfitting while increasing the scope of the approach: Gaussian base functions in addition to standard sigmoids, adaptive selection of the type and number of base functions \cite{Belyaev2013}, a new algorithm of initialization of base functions' parameters~\cite{Burnaev2016b}, adaptive regularization~\cite{Belyaev2013}, boosting used to construct ensembles for additional improvement of accuracy and stability~\cite{Burnaev2013}, post-processing of the results to remove redundant features. HDAGP~\cite{Burnaev2016} extends GP by adding to it HDA-based non-stationary covariance functions with the goal of improving GP's ability to deal with spatially inhomogeneous dependencies.

\subsubsection*{Tensor Products of Approximations (TA), incomplete Tensored Approximations (iTA), and Tensored Gaussian Processes (TGP)} 
\begin{figure}
	\centering
    \includegraphics[width=0.5\textwidth]{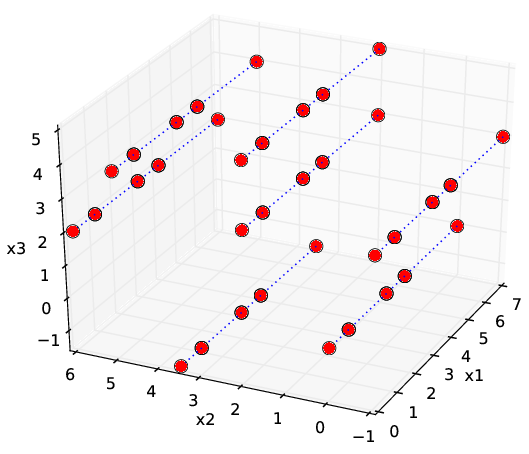}
    \caption{Example of factorization of a three-dimensional DoE consisting of 35 points: the DoE is a Cartesian product of its two-dimensional projection to the $x_2x_3$-plane (of size 7) with its one-dimensional projection to the $x_1$-axis (of size 5). The dotted lines connect points with the same projection to the $x_2x_3$-plane.} 
    \label{fig:factored_doe}
\end{figure}

TA \cite{Belyaev2015b} is not a single approximation method, but rather a general and very flexible construction addressing the issue of anisotropy mentioned in the introduction. In a nutshell, TA is about forming spatial products of different approximation techniques, with each technique associated with its own subset of input variables. The key condition under which TA is applicable is the factorizability of the DoE: the DoE must be a Cartesian product of some sets with respect to some partition of the whole collection of input variables into sub-collections, see a two-factor example in Figure \ref{fig:factored_doe}. If TA is enabled, GTApprox automatically finds the most detailed factorization for the DoE of the given training set. Once a factorization is found, to each factor one can assign a suitable approximation technique, and then form the whole approximation using a ``product'' of these techniques. This essentially means that the overall approximation's dictionary of basis functions is formed as the product of the factors' dictionaries. The coefficients of the expansion over this dictionary can be found very efficiently \cite{Belyaev2015b,Belyaev2015c}.

\begin{figure}
	\centering
    \begin{subfigure}[b]{0.45\textwidth}
    \centering
    \includegraphics[width=\textwidth, clip, trim=10mm 0mm 10mm 10mm]{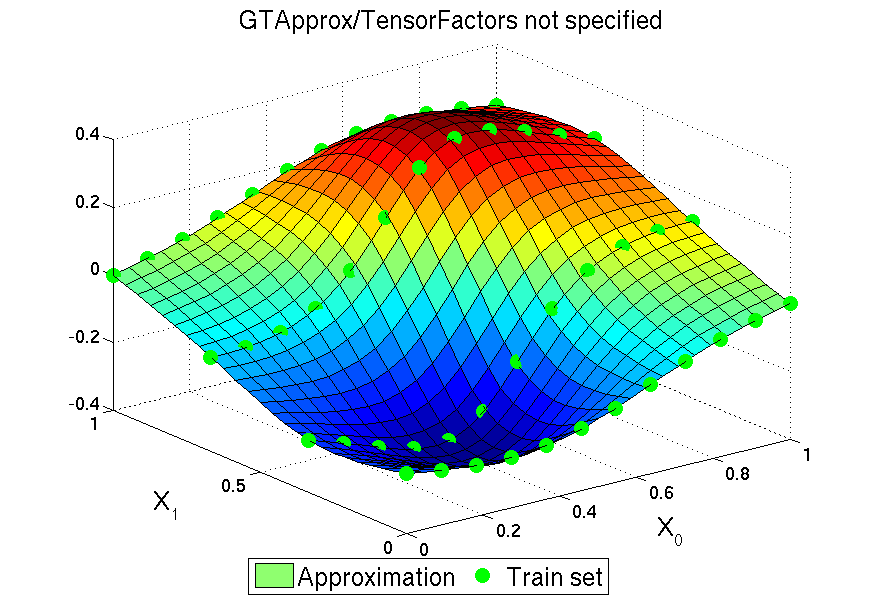}
    \caption{}
    \end{subfigure}
    \begin{subfigure}[b]{0.45\textwidth}
    \centering
    \includegraphics[width=\textwidth, clip, trim=10mm 0mm 10mm 10mm]{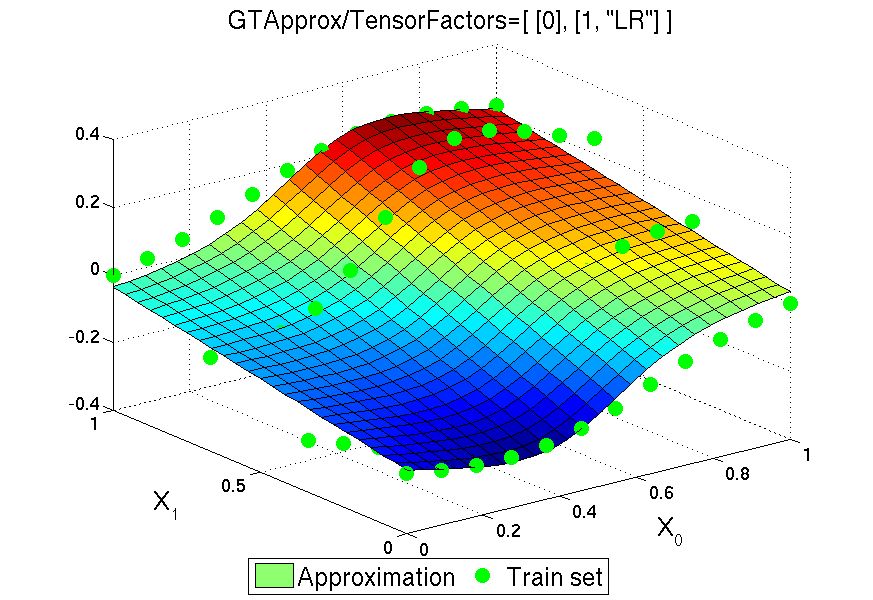}
    \caption{}
    \end{subfigure}
    \caption{Two TA approximations for the same training set with two one-dimensional factors: (a) the default approximation using splines  in both factors; (b) an approximation with splines in the first factor and Linear Regression in the second factor. Note that approximation (b) depends linearly on $x_1$.} 
    \label{fig:TAexample}
\end{figure}

GTApprox offers a number of possible techniques that can be assigned to a factor, including Linear Regression (LR), B-splines (BSPL), GP and HDA, see example in Figure \ref{fig:TAexample}. It is natural, for example, to assign BSPL to one-dimensional factors and LR, GP or HDA to multi-dimensional ones. If not assigned by the user, GTApprox automatically assigns a technique to each factor by a heuristic akin to the decision tree described later in section \ref{sec:algo_sel}.  

Factorizability of the DoE is not uncommon in engineering practice. Note, in particular, that the usual full-factorial DoE is a special case of factorizable DoE with one-dimensional factors. Also, factorization often occurs in various scenarios where some input variables describe spatial or temporal location of measurements within one experiment while other variables describe external conditions or parameters of the experiment -- in this case the two groups of variables are typically varied independently. Moreover, there is often a significant anisotropy of the DoE with respect to this partition: each experiment can be expensive, but once the experiment is performed the values of the monitored quantity can be read off of multiple locations relatively easily, so the DoE factor associated with locations is much larger than the factor associated with experiment's parameters. The advantage of TA is that it can overcome this anisotropy by assigning to each DoE factor a separate technique, most appropriate for this particular factor. 

\begin{figure}
	\centering
    \begin{subfigure}[b]{0.45\textwidth}
    \centering
    \includegraphics[width=\textwidth, clip, trim=60mm 30mm 20mm 15mm]{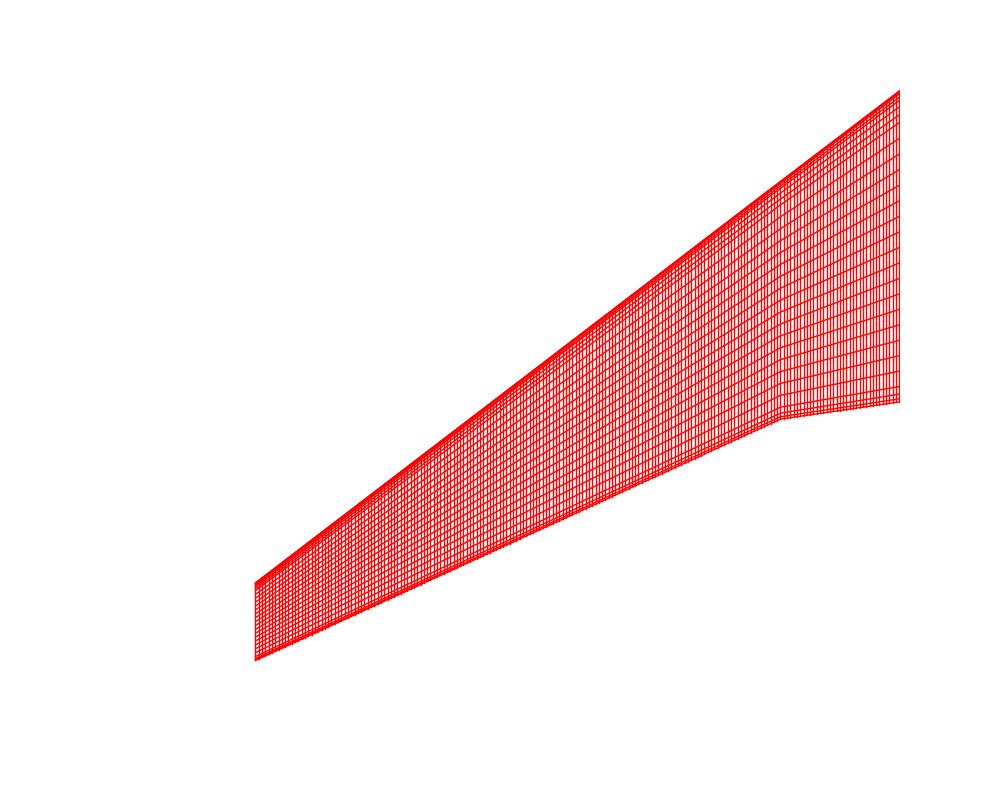}
    \caption{}
    \end{subfigure}
    \begin{subfigure}[b]{0.45\textwidth}
    \centering
    \includegraphics[width=\textwidth, clip, trim=43mm 15mm 38mm 10mm]{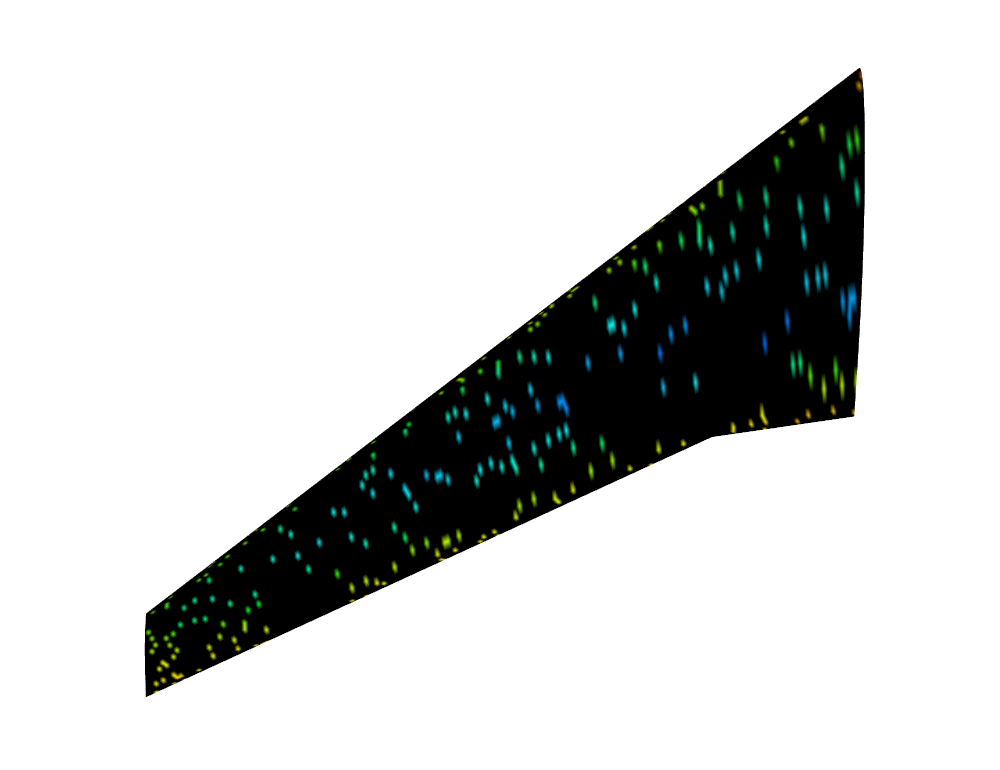}
    \caption{}
    \end{subfigure}
     \begin{subfigure}[b]{0.9\textwidth}
    \centering
    \includegraphics[width=\textwidth, clip, trim=115mm 95mm 115mm 80mm]{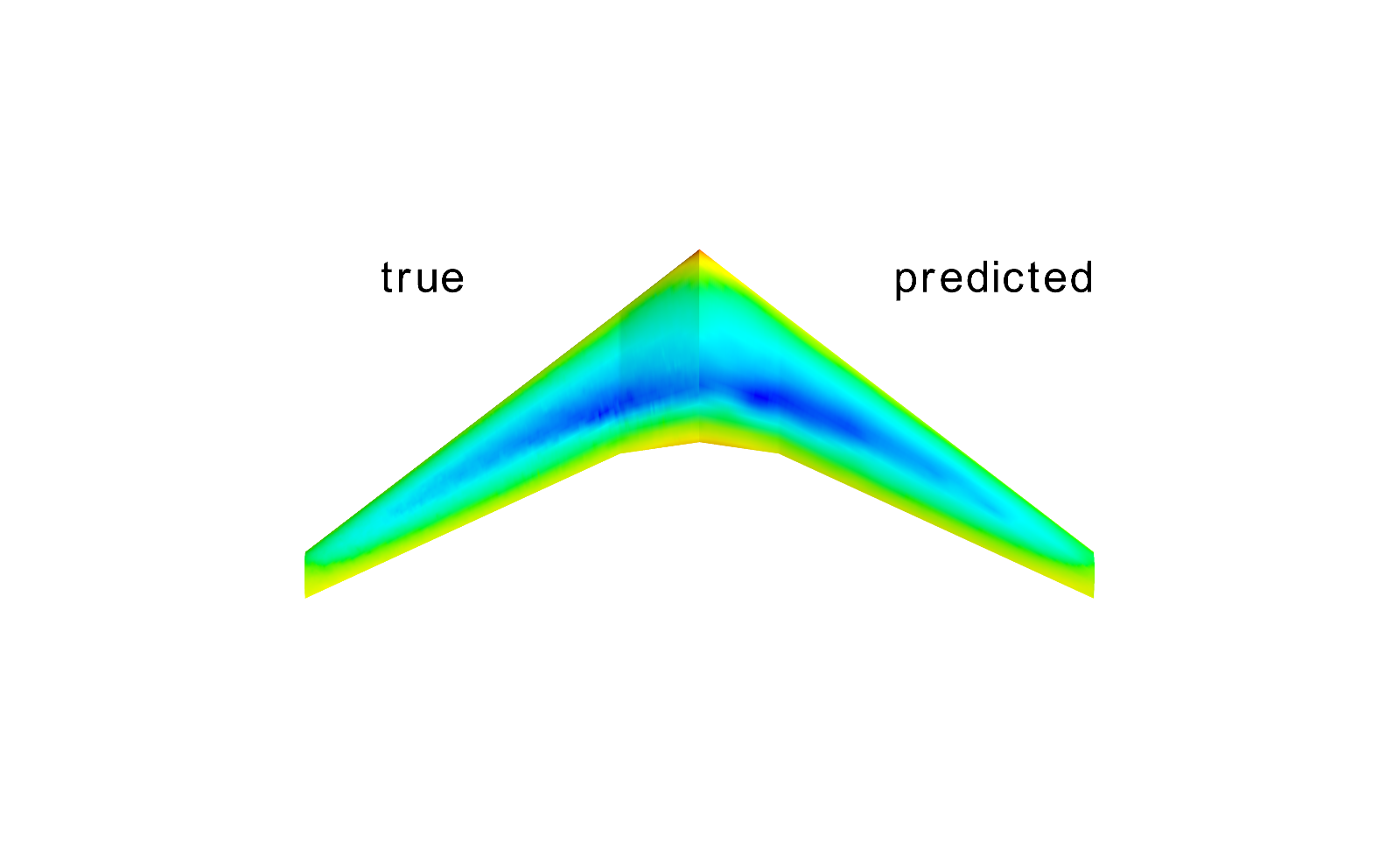}
    \caption{}
    \end{subfigure}
    \caption{Application of iTA to reconstruction of a pressure distribution on a wing. The distribution was obtained by an aerodynamic simulation. (a) The $200\times 29$ grid on which the pressure distribution is defined. (b) iTA is applied to a training set of pressure values at 290 randomly chosen points. (c) The resulting approximation is compared with the true distribution.} 
    \label{fig:wing}
\end{figure}

Nevertheless, exact factorizability is, of course, a relatively restrictive assumption. The incomplete Tensored Approximation (iTA) technique of GTApprox relaxes this assumption: this technique is applicable if the DoE is only a \emph{subset} of a full-factorial DoE and all factors are one-dimensional. This covers a number of important use cases: a full-factorial DoE where some experiments are not finished or the solver failed to converge, or a union of several full-factorial DoEs resulting from different series of experiments, or a Latin Hypercube on a grid. Despite the lack of Cartesian structure, construction of the approximation in this case reduces to a convex quadratic programming problem leading to a fast and accurate solution~\cite{Belyaev2015c}. An example of iTA's application to a pressure distribution on a wing is shown in Figure \ref{fig:wing}. Also, see Section~\ref{sec:tensor} for an industrial example.  

Tensored Gaussian Processes (TGP)~\cite{Belyaev2015,Belyaev2016} is yet another incarnation of tensored approximations. TGP is fast and intended for factorized DoE like the TA technique, but is equipped with local accuracy estimates like GP.

\begin{figure}
	\centering
    \begin{subfigure}[b]{0.45\textwidth}
    \centering
    \includegraphics[width=\textwidth, clip, trim=0mm 20mm 0mm 0mm]{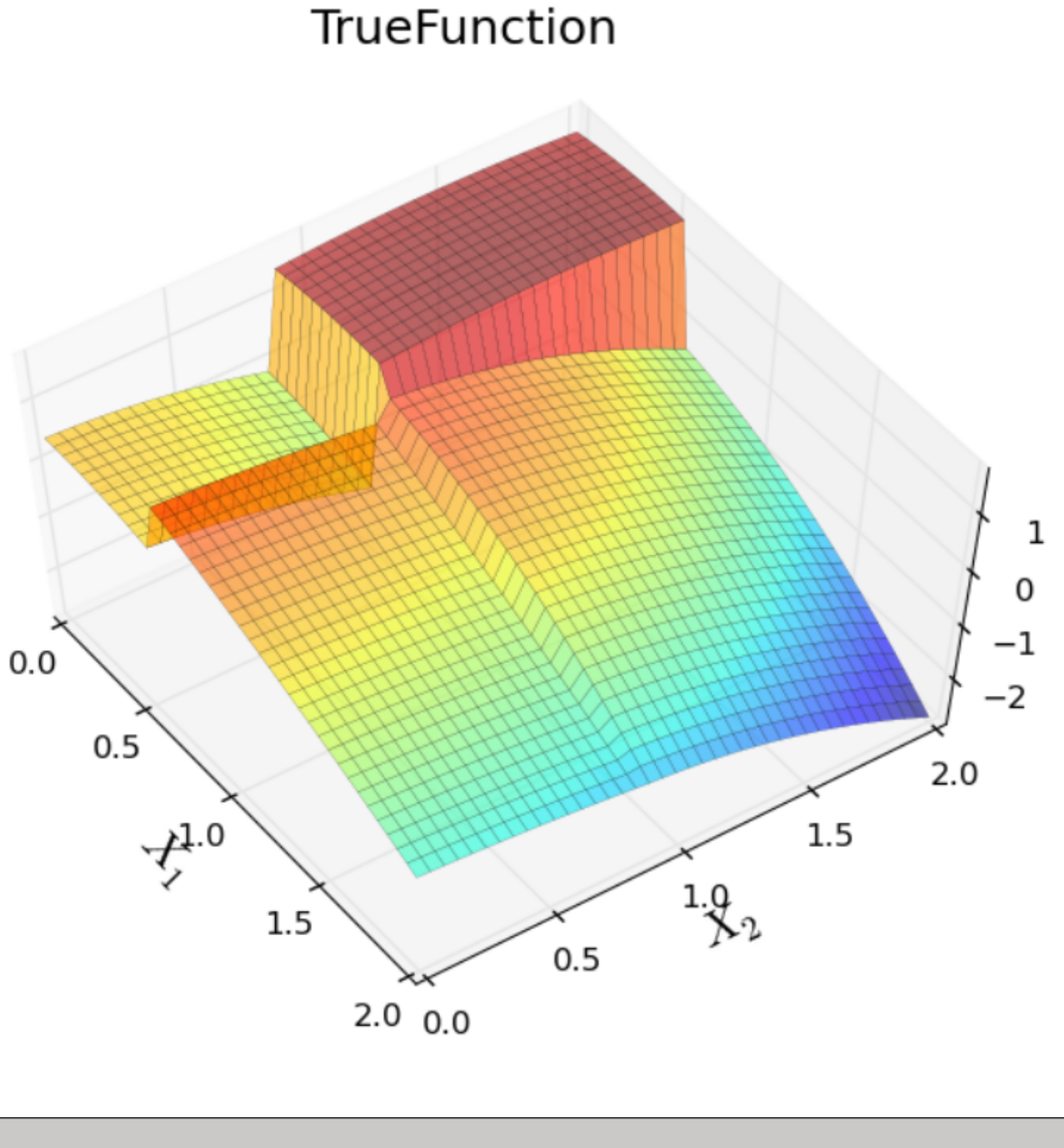}
    \caption{}
    \end{subfigure}
    \begin{subfigure}[b]{0.9\textwidth}
    \centering
    \includegraphics[width=\textwidth, clip, trim=0mm 0mm 0mm 0mm]{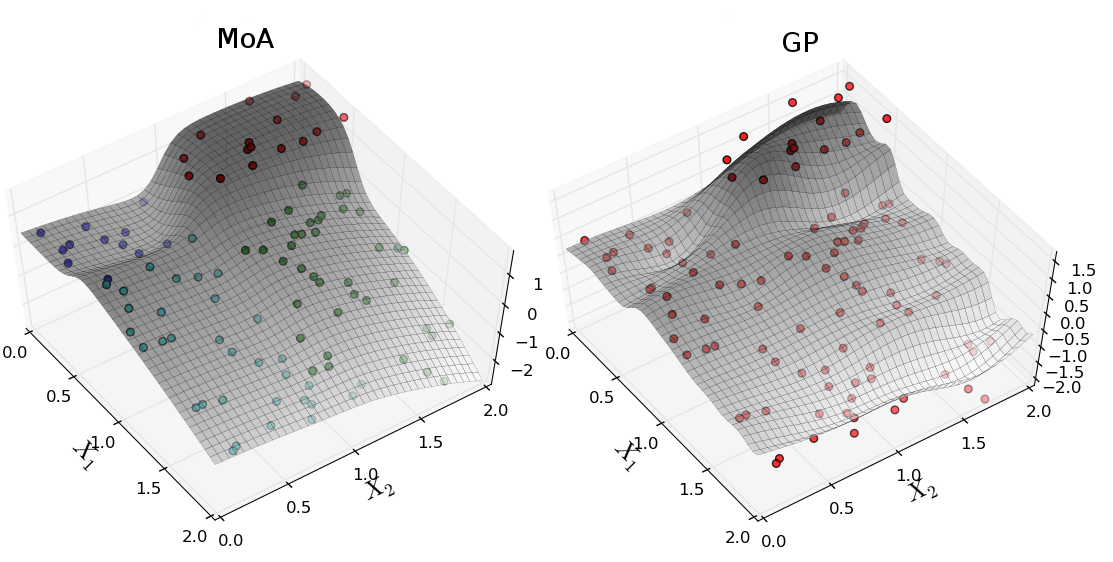}
    \caption{}
    \end{subfigure}
    \caption{Application of MoA to a spatially inhomogeneous response function. (a) The true response function. (b) Approximations by MoA and GP.} 
    \label{fig:samgen}
\end{figure}

\paragraph*{Mixture of Approximations (MoA)}
If the response function is very inhomogeneous, a single surrogate model may not efficiently cover the whole design space (see \cite{Grihon2013} and Section~\ref{sec:copti} for an engineering example with critical buckling modes in composite panels).
One natural approach to overcome this issue is to perform a preliminary space partitioning and then build a separate model for each part. This is exactly what Mixture of Approximations does. This technique falls into the family of Hierarchical Mixture Models \cite{Jordan1994,Yuksel2012}. A Gaussian mixture model is used to do the partitioning, and after that other techniques are used to build local models. MoA is implemented to automatically estimate the number of parts; it supports possibly overlapping parts and preservation of model continuity across different parts~\cite{Grihon2013}. A comparison of MoA with a standard technique (GP) is shown in  Figure~\ref{fig:samgen}. 

\paragraph*{Gradient Boosted Regression Trees (GBRT)} This is a well-known technique that uses decision trees as weak estimators and combines several weak estimators into a single model, in a stage-wise fashion~\cite{Friedman2001}. 
GBRT is suitable for problems with large data sets and in cases when smooth approximation is not required.

\paragraph*{Piece-wise Linear Approximation (PLA)} This technique is based on the Delaunay triangulation of the training sample. It connects neighboring points of the given training set into triangles (or tetrahedrons) and builds a linear model in each triangle. 
PLA is a simple and reliable interpolation technique. It is suitable for low-dimensional problems (mostly 1D, 2D and 3D) where the approximation is not required to be smooth. In higher dimensions the construction becomes computationally intractable. 

\subsection{User options and additional features}

GTApprox implements a number of user options and additional features that directly address the issues raised in the Section~\ref{section: introduction}. The options are not linked to specific approximation techniques described in the previous subsection; rather, the tool selects and tunes a technique according to the options (see Section \ref{sec:algo_sel}). The formulations of options and features avoid references to algorithms' details; rather, they are described by their overall effect on the surrogate model. Below we list a few of these options and features.

\paragraph*{Accelerator} Some techniques contain parameters significantly affecting the training time of the surrogate model (e.g., the number of basic approximators in HDA or the number of decision trees in GBRT). By default, GTApprox favors accuracy over training time. The Accelerator option defines a number of ``levels''; each level assigns to each technique a set of parameters ensuring that the training time with this technique is increasingly reduced as the level is increased.   

\paragraph*{Gradient/Jacobian matrix} To serve optimization needs, each approximation produced by GTApprox (except non-smooth models, i.e. GBRT and PLA) is constructed simultaneously with its gradient (or Jacobian matrix in the context of multi-component approximations).

\paragraph*{Accuracy Evaluation (AE)} Some GTAppox' techniques of Bayesian nature (mostly GP-based, i.e. GP, SGP, HDAGP, TGP) construct surrogate models along with point-wise estimates of deviations of these models from true response values~\cite{Rasmussen2005}. These estimates can be used in SBO, to define an objective function taking into account local uncertainty of the current approximation (see an example in Section \ref{sec:hyper_sel}). 

\paragraph*{Smoothing} Each approximation constructed by GTApprox can be additionally smoothed. The smoothing is done by additional regularization of the model; the exact algorithms depends on the particular technique. Smoothing affects the gradient of the model as well as the model itself. Smoothing may be useful in tasks for which smooth derivatives are important, e.g., in surrogate-based optimization.

\paragraph*{Componentwise vs. joint approximation} If the response function has several scalar components, approximation of all components one-by-one can be lengthy and does not take into account relations that may connect different components. Most of the GTApprox' techniques have a special ``joint'' mode where the most computationally intensive steps like iterative optimization of the basis functions in HDA or kernel optimization in GP is performed only once, simultaneously for all output components, and only the last step of linear expansion over basis functions is performed separately for each output~\cite{Burnaev2016}. This approach can significantly speed up training. For example, training of a GP model with $m$ outputs with a training set of $N$ points requires $O(m N^3)$ arithmetic operations in the componentwise mode, while in the joint mode it is just $O(N^3 + m N^2)$. Furthermore, because of the partly shared approximation workflow, the joint mode better preserves similarities between different components of the response function (whenever they exist).

\paragraph*{Exact Fit} Some GTApprox' techniques (like GP and splines) allow to construct approximations that pass exactly through the points of the training set. Note, however, that this requirement may sometimes lead to overfitting and is certainly not appropriate if the training set is noisy. 

\section{Automated choice of the technique}\label{sec:algo_sel}
GTApprox implements two meta-algorithms automating the choice of the approximation technique for the given problem: a simpler one, based on hand-crafted rules and hereafter referred to as the ``Decision Tree'', and a more complex one, including problem-specific adaptation and branded as ``Smart Selection''. We outline these two meta-algorithms below.

\subsection{``Decision tree''} \label{sec:dectree}
\begin{figure}
\begin{center}
\includegraphics[scale=0.3]{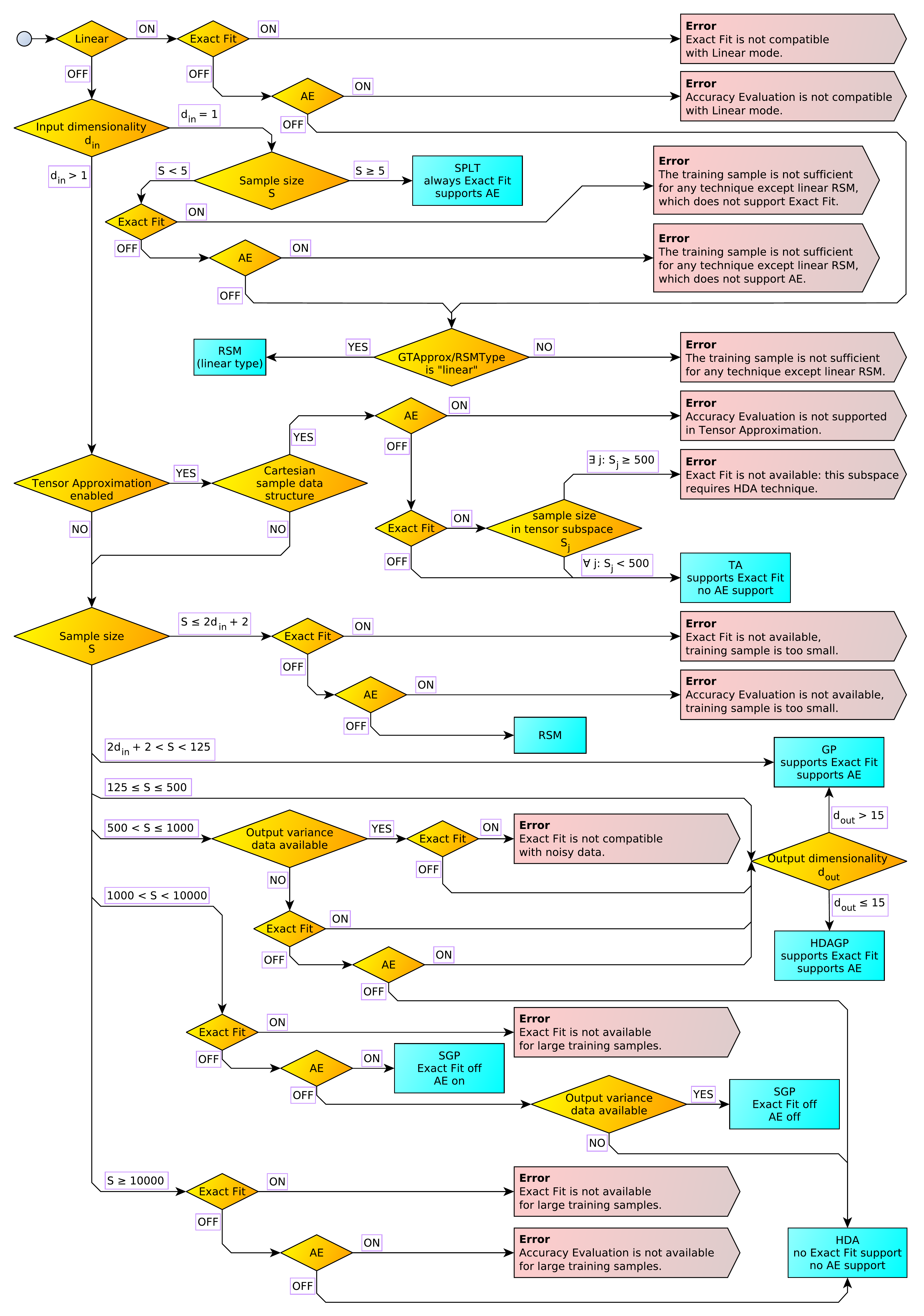}
\end{center}
\caption{The ``decision tree'' technique selection in GTApprox. Rectangles show individual techniques. Rhombuses show choices depending on properties of the training data and user options. Pentagons show exceptional cases with conflicting or unfeasible requirements.}\label{fig:dec_tree}
\end{figure}

The ``decision tree'' approach selects an appropriate technique using predetermined rules that involve size and dimensions of the data sets along with user-specified features (requirements of model linearity, Exact Fit or Accuracy Evaluation, enabled Tensor Approximations), see Figure \ref{fig:dec_tree}. The rules are partly based on the factual capabilities and limitations of different techniques and partly on extensive preliminary testing and practical experience. The rules do not guarantee the optimal choice of a technique, but rather select the most reasonable candidate.  

\subsection{``Smart selection''}\label{sec:hyper_sel}
The main drawbacks of the ``decision tree'' approach are that it only takes into account the crudest properties of the data set (size, dimensions) and cannot adjust parameters of the technique, which is often important.   

To address both issues, ``Smart selection'' performs, for each training set, a numerical optimization of the technique as well as its parameters \cite{Snoek2012,Bergstra2011}, by minimizing the cross-validation error.

This is a quite complex optimization problem: the search space is tree-structured, parameters can be continuous or categorical, the objective function is noisy and expensive to evaluate.

We first describe how we optimize the vector of parameters for a given technique. To this end we use Surrogate Based Optimization (SBO) \cite{Jones1998,Brochu2010}.
Recall that SBO is an iterative algorithm that can be written as follows:

\begin{enumerate}
    \item Pick somehow an initial candidate $\boldsymbol{\lambda}_1$ for the optimal vector of parameters.  
    \item Given the current candidate $\boldsymbol{\lambda}_k$ for the optimal vector of parameters, find the value $c_k$ of the objective function (cross-validation error) on it.
    \item Using all currently available pairs $R = (\boldsymbol{\lambda}_i, c_i)_{i = 1}^k$, construct an \emph{acquisition function} $a(\boldsymbol{\lambda}; R)$ reflecting our preference for $\boldsymbol{\lambda}$ to be the next candidate vector.
    \item Choose the new candidate vector of parameters $\boldsymbol{\lambda}_{k+1}$ by numerically optimizing the acquisition function and return to step 2.
\end{enumerate}

The acquisition function used in step 3 must be a reasonably lightweight function involving not only the current estimate of the objective function $c(\boldsymbol{\lambda})$, but also the uncertainty of this estimate, in order to make incentive for the algorithm to explore new regions of the parameter space. A standard choice for the acquisition function that we use is the {\em Expected Improvement} function
\[
    a_{EI}(\boldsymbol{\lambda}; R) = \mathbb{E}((c' - c(\boldsymbol{\lambda}))_+),
\]
where $c'=\min_{1\le i\le k}{c_i}$ is the currently known minimum and $(c' - c(\boldsymbol{\lambda}))_+=\max(0,c' - c(\boldsymbol{\lambda}))$ is the objective function's improvement resulting from considering a new vector $\boldsymbol{\lambda}$. The expectation here can be approximately written, under assumption of a univariate normal distribution of error, in terms of the expected value $\widehat{c}(\boldsymbol{\lambda})$ of $c(\boldsymbol{\lambda})$ and the expected value $\widehat{\sigma}(\boldsymbol{\lambda})$ of the deviation of $c(\boldsymbol{\lambda})$ from $\widehat{c}(\boldsymbol{\lambda})$. The function $\widehat{c}(\boldsymbol{\lambda})$ is found as a GTApprox surrogate model constructed from the data set $R$, and the accompanying uncertainty estimate $\widehat{\sigma}(\boldsymbol{\lambda})$ is found using the Accuracy Evaluation feature.

The described procedure allows us to choose optimal parameters for a particular technique.
In order to choose the technique we perform SBO for each technique from some predefined set, and then select the technique with the minimal error.

The set of techniques is formed according to {\em hints} specified by the user. The hints are a generalization of options towards less technical and more intuitive description of data or of the required properties of the surrogate model. In general, hints may be imprecise, e.g. ``IsNoisy'' or ``ClusteredData''. Hints may play the role of tags or keywords helping the users to express their domain-specific knowledge and serving to limit the range of techniques considered in the optimization. 

The ``smart selection'' approach is time consuming, since each SBO iteration involves constructing a new auxiliary surrogate model. The process can be sped up by a few hints. The ``Accelerator'' hint adjusts parameters of the SBO procedure, making it faster but less accurate. The ``AcceptableQualityLevel'' hint allows the user to specify an acceptable level of model accuracy for an early stopping of SBO.

There exist general-purpose Python frameworks for optimizing parameters of approximation techniques, e.g. hyperopt \cite{Bergstra2013}. We have considered using hyperopt (via HPOlib, \cite{hpolib}) with GTApprox as an alternative to ``smart selection'', but found the results to be worse than with ``smart selection''. First, being a general framework, HPOlib/hyperopt does not take into account special properties of particular techniques. For example, GP-based techniques have high computational complexity and cannot be applied in the case of large training sets, but HPOlib/hyperopt would attempt to build a GP model anyway. Second, the only termination criterion in HPOlib/hyperopt is the maximum number of constructed models -- a criterion not very flexible given that different models can have very different training times. Finally, we have observed HPOlib/hyperopt in some cases to repeatedly construct models with the same parameters, which is again inefficient since training times for some of our models are quite large.  

\section{Comparison with alternative algorithms on test problems}\label{sec:tests}

We perform a comparison of accuracy between GTApprox and some of the most popular, state-of-the-art predictive modeling Python libraries: scikit-learn \cite{Sklearn2011}, XGBoost \cite{xgboost}, and GPy \cite{gpy}.\footnote{The code and data for this benchmark are available at \url{https://github.com/yarotsky/gtapprox_benchmark}. The versions of the libraries used in the benchmark were GTApprox 6.8, scikit-learn 0.17.1, XGBoost 0.4, and GPy 1.0.9.} We emphasize that there are a few caveats to this comparison. First, these libraries are aimed at a technically advanced audience of data analysts who are expected to themselves select appropriate algorithms and tune their parameters. In particular, scikit-learn does not provide a single entry point wrapping multiple techniques like GTApprox does, as described in Section \ref{sec:algo_sel}. We therefore compare GTApprox, as a single algorithm, to multiple algorithms of scikit-learn. We also select a couple of different modes in both XGBoost and GPy. Second, the scope of scikit-learn and XGBoost is somewhat different from that of GTApprox: the former are not focused on regression problems and their engineering applications and, in particular, their most powerful nonlinear regression methods seem to be ensembles of trees (Random Forests and Gradient Boosted Trees) that produce piece-wise constant approximations presumably not fully suitable for modeling continuous response functions. Keeping these points in mind, our comparison should be otherwise reasonably fair and informative.

We describe now the specific techniques considered in the benchmark. All techniques are used with default settings.  

We consider a diverse set of scikit-learn methods for regression, both linear and nonlinear:  Ridge Regression with cross-validation (denoted by \textsf{SL\_RidgeCV} in our tests), Support Vector Regression (\textsf{SL\_SVR}), Gaussian Processes (\textsf{SL\_GP}), Kernel Ridge (\textsf{SL\_KR}), and Random Forest Regression (\textsf{SL\_RFR}). Our preliminary experiments included more methods, in particular common Linear Regression, LassoCV and Gradient Boosting, but we have found their results to be very close to results of other linear or tree-based methods. 

We consider two modes of XGBoost: with the {\tt gbtree} booster (default, \textsf{XGB}) and with the {\tt gblinear} booster (\textsf{XGB\_lin}).

We consider two modes of GPy: the {\tt GPRegression} model (\textsf{GPy}) and, since some of our test sets are relatively large, the {\tt SparseGPRegression} model (\textsf{GPy\_sparse}).

Finally, we consider two versions of GTApprox corresponding to the two meta-algorithms described in Section \ref{sec:algo}: the basic tree-based algorithm (\textsf{gtapprox}) and the ``smart selection'' algorithm (\textsf{gta\_smart}).

Our test suite contains 31 small- and medium-scale problems, of which 23 are given by explicit formulas and the remaining 8 represent real-world data sets or results of complex simulations. The problems defined by formulas include a number of functions often used for testing optimization algorithms~\cite{OptTestFunctions}, such as Ackley function, Rosenbrock function, etc. Additionally, they include a number of non-smooth and noisy functions. The real-world data sets and data of complex simulations are borrowed from the UCI repository \cite{Lichman:2013} and the GdR Mascot-Num benchmark \cite{MascotNum}. Detailed descriptions or references for the test problems can be found in the GTApprox documentation (\cite[MACROS User Manual, section ``Benchmarks and Tests'']{datadvance_docs}). 

Each problem gives rise to several tests by varying the size of the training set and the way the training set is generated: for problems defined by explicit functions we create the training set by evaluating the response function on a random DoE or on a Latin Hypercube DoE of the given size; for problems with already provided data sets we randomly choose a subset of the given size. As a result, the size of the training set in our experiments ranges from 5 to 30000. Testing is performed on a holdout set. Some of the problems have multi-dimensional outputs; in such cases each scalar output is handled independently and is counted as a separate test. The total number of tests obtained in this way is 430. Input dimensionality of the problems ranges from 1 to 20.   

In each test we compute the relative root-mean-squared prediction error as
\begin{equation*}
\mathrm{RRMS} = \Bigg(\frac{\sum_{n=1}^M \big(f(\mathbf{x}_n) -\widehat{f}(\mathbf{x}_n)\big)^2}{\sum_{n=1}^M \big(f(\mathbf{x}_n) -\overline{f}\big)^2}\Bigg)^{1/2},
\end{equation*}
where $\big(\mathbf{x}_n, f(\mathbf{x}_n)\big)_{n=1}^M$ is the test set with true values of the response function, $\widehat{f}(\mathbf{x}_n)$ is the predicted value, and $\overline{f}$ is the mean value of $f$ on the test set. Note that RRMS essentially compares surrogate models $\widehat{f}$ with the trivial constant prediction $\overline{f}$, up to the fact that $\overline{f}$ is computed on the test set rather than the training set.  

Each test is run in a separate OS process with available virtual memory restricted to 6GB. Some of the techniques raise exceptions when training on certain problems (e.g., out-of-memory errors). In such cases we set $\mathrm{RRMS} = +\infty$. 

For each surrogate modeling algorithm we construct its accuracy profile as the function showing for any RRMS threshold the ratio of tests where the RRMS error was below this threshold. 

\begin{figure}
    \centering
    \includegraphics[width=0.7\textwidth, clip, trim=10mm 0mm 15mm 10mm]{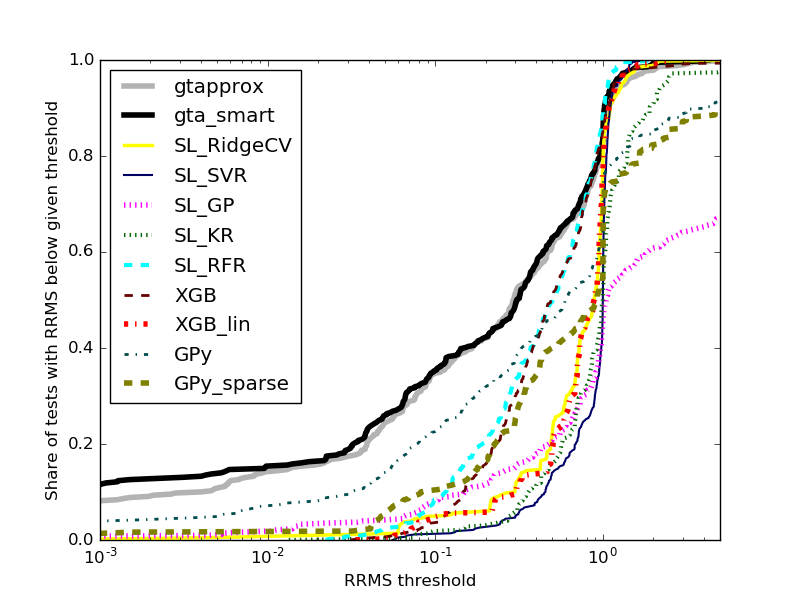}
    \caption{Accuracy profiles of different approximation algorithms}
    \label{fig:accuracy_profiles}
\end{figure}

The resulting profiles are shown in Figure \ref{fig:accuracy_profiles}. We find that, on the whole, the default GTApprox is much more accurate than default implementations of methods from other libraries, with the exception of highly noisy problems where $\mathrm{RRMS}>1$: here \textsf{gtapprox} performs just a little worse than linear and tree-based methods. As expected, \textsf{gta\_smart} yields even better results than \textsf{gtapprox}.

\begin{figure}
    \centering
    \includegraphics[width=0.7\textwidth, clip, trim=10mm 0mm 15mm 10mm]{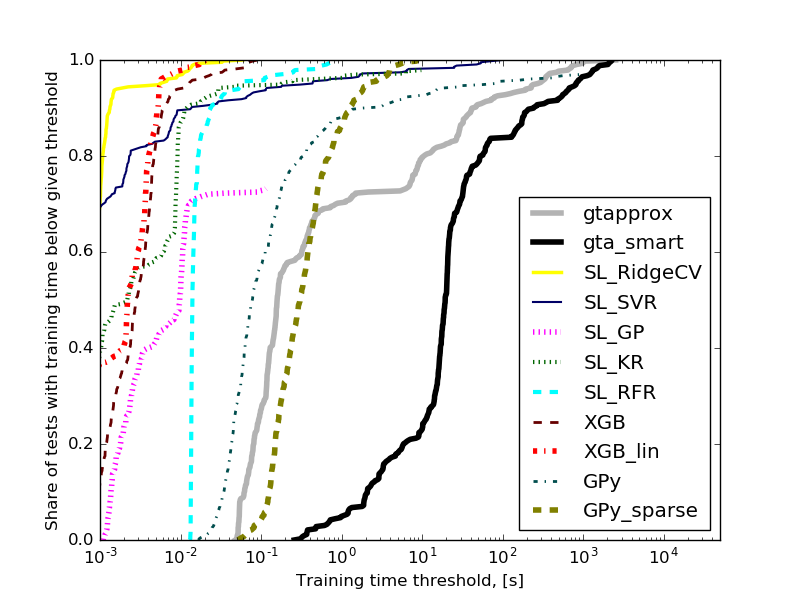}
    \caption{Training time profiles of different approximation algorithms}
    \label{fig:time_profiles}
\end{figure}

We should, however, point out that this advantage in accuracy is achieved at the cost of longer training. Possibly in contrast to other tools, GTApprox favors accuracy over training time, assuming the user of the default algorithm delegates to it the experiments needed to obtain an accurate model. In Figure \ref{fig:time_profiles} we show profiles for training time. Whereas training of scikit-learn and XGBoost algorithms with default settings typically takes a fraction of second, GTApprox may need a few minutes, especially the ``smart selection'' version. Of course, if desired, training time can be reduced (possibly at the cost of accuracy) by tuning various options of GTApprox.   

\section{Applications}\label{sec:apps}

We briefly describe several industrial applications of GTApprox \cite{Grihon2013, Grihon2014, Struzik2013, Belyaev2014} to illustrate how special features of GTApprox can help in solving real world problems.

\subsection{Surrogate models for reserve factors of composite stiffened panels}
\label{sec:copti}

\begin{figure}[!tbp]
    \centering
    \includegraphics[width=0.3\textwidth, clip, trim=0mm 0mm 0mm 0mm]{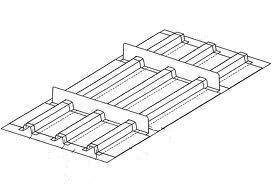}
    \caption{A stiffened panel.}
    \label{fig:copti_stringer}
\end{figure}

Aeronautical structures are mostly made of stiffened panels that consist of thin shells (or skins) enforced with stiffeners in two orthogonal directions (see Figure~\ref{fig:copti_stringer}). The stiffened panels are subject to highly nonlinear phenomena such as buckling or collapse. Most strength conditions for the structure's reliability can be formulated using so-called reserve factors (RFs). In the simplest case, a reserve factor is the ratio between an allowable stress (for example, material strength) and the applied stress. The whole structure is validated if all RFs of all the panels it consists of are greater than 1. RF values are usually found using computationally expensive Finite Element (FE) methods. 

During the sizing process, i.e. optimizing geometry of the structure with respect to certain criteria (usually minimization of the weight of the structure), RF values are taken as optimization constraints that allow to conclude if the considered geometry would be reliable. So for all basic structures the RFs and their gradients have to be recomputed on every optimization step, which becomes a very expensive operation in terms of time.

The goal of this application was to create a  surrogate model that works orders of magnitude faster than the FE method and at the same time has a good accuracy: error should be less than 5\% for at least 95\% of points with RF values close to 1, and the model should reliably tell if the RF is greater or less than 1 for a particular design.

The complexity of the problem was exacerbated by several issues. First, the RF values depend on 20 parameters (geometry and loads), all of which significantly affect the output values. Second, some RFs depend on the parameters discontinuously. Third, points with RFs close to 1 are scattered across the input domain.

\begin{figure}
    \centering
    \includegraphics[width=0.9\textwidth, clip, trim=0mm 0mm 0mm 0mm]{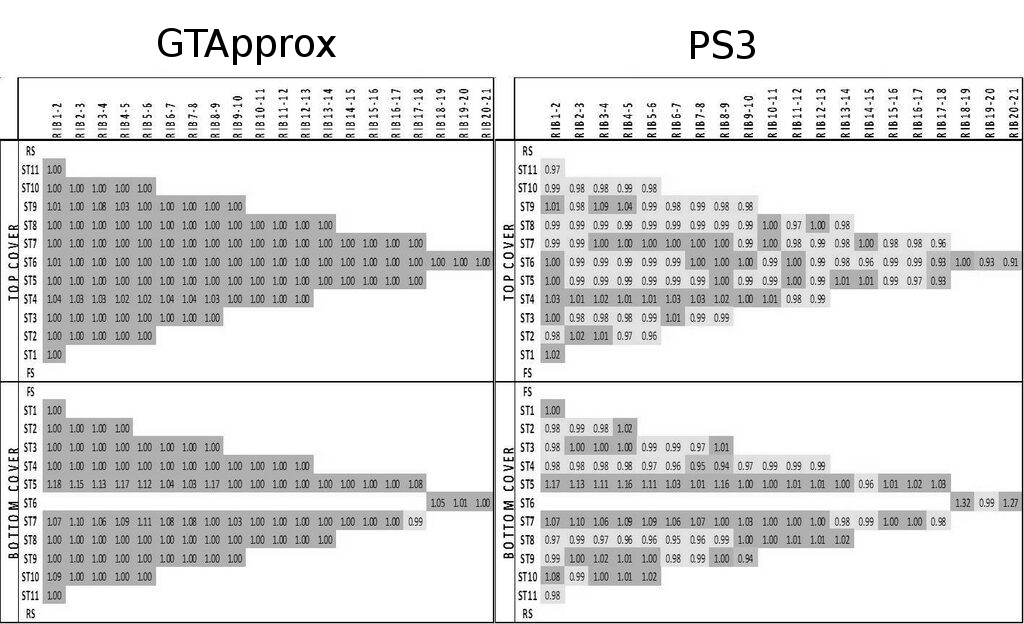}
    \caption{RF values at the optimum. Comparison of the optimal result found using the GTApprox surrogate model with that of PS3.}
    \label{fig:copti_example}
\end{figure}

The Mixture of Approximation (MoA) technique of GTApprox was used to create a surrogate model based on the train dataset of 200000 points that met the accuracy requirements and worked significantly faster than the reference PS3 tool implementing the FE computation. The optimization was further facilitated by the availability of the gradients of the GT Approx model. 
The optimization results obtained by GTApprox and the PS3 tool are shown in  Figure~\ref{fig:copti_example}. Details on the work can be found in~\cite{Grihon2013, Grihon2014}. The obtained GTApprox surrogate model was embedded into the pre-sizing optimization process of Airbus A350XWB composite boxes.

\subsection{Surrogate models for helicopter loads estimation} 
\label{sec:helicopter}

In this work GTApprox was used to create surrogate models for maximum loads acting on various structural elements of the helicopter. Knowledge of maximum loads during the flight allows one to estimate fatigue and thus see when repair is needed. Such data can only be collected in the flight tests as one needs to install additional sensors to a helicopter to measure loads, which are too expensive to be installed on every machine.

So the goal of the project was to take data already measured during flight tests and create surrogate models that would allow to estimate loads on every flight as a function of flight parameters and external conditions. The challenge of the project was that models for lots of different load types and flight conditions (e.g. maneuver types) needed to be created. In total one needed to build $4152$ surrogate models. Such problem scale made it impossible to tune each model ``manually''. And at the same time different combinations of loads and flight condition could demonstrate very different behavior and depend on different set of input parameters. The input dimension varied in the range from 8 to 10 and the sample size was from 1 to 108 points.

GTApprox' capabilities on automatic technique selection and quality assessment were used to create all $4152$ models with the required accuracy without manually tweaking their parameters in each case. In total, $2877$ constant models, $777$ RSM models, $440$ GP models and $58$ HDA models were constructed. Only a few most complex cases had to be specifically addressed in an individual manner. More details on the work can be found in~\cite{Struzik2013}.

\subsection{Surrogate models for aerodynamic problems} 
\label{sec:tensor}
In this application GTApprox was used to obtain surrogate models for aerodynamic response functions of 3-dimensional flight configurations~\cite{Belyaev2014}. The training data were obtained either by Euler/RANS CFD simulations or by wind tunnel tests; in either case experiments were costly and/or time-consuming, so a surrogate model was required to cover the whole domain of interest.

\begin{figure}
    \centering
    \includegraphics[width=0.6\textwidth, clip, trim=20mm 5mm 40mm 20mm]{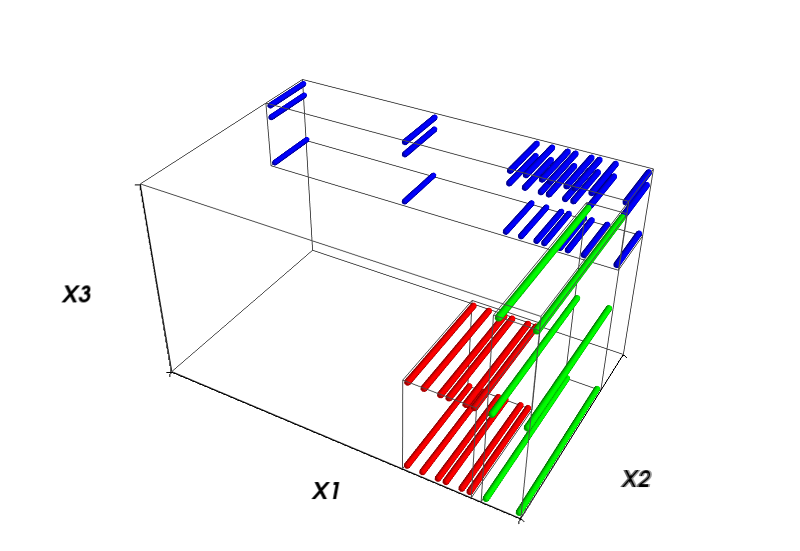}
    \caption{Design of Experiments in the aerodynamic test case.}
    \label{fig:doe_ita}
\end{figure}

The training set's DoE, shown in Figure \ref{fig:doe_ita}, had two important peculiarities. First, the DoE was a union of several (irregular) grids resulting from different experiments. Second, the grids were highly anisotropic:  variables $x_1, x_3$ were sampled with much lower resolutions than variable $x_2$.    

As explained in Section \ref{sec:algo}, this complex structure is exactly what the iTA technique is aimed at. The whole DoE can be considered as an incomplete grid, so iTA is directly applicable to the whole training set without the need to construct and then merge separate approximations for different parts of the design space. 

\begin{figure}
    \centering
    \begin{subfigure}[b]{0.48\textwidth}
      \includegraphics[width=1.0\textwidth, clip, trim=60mm 0mm 70mm 10mm]{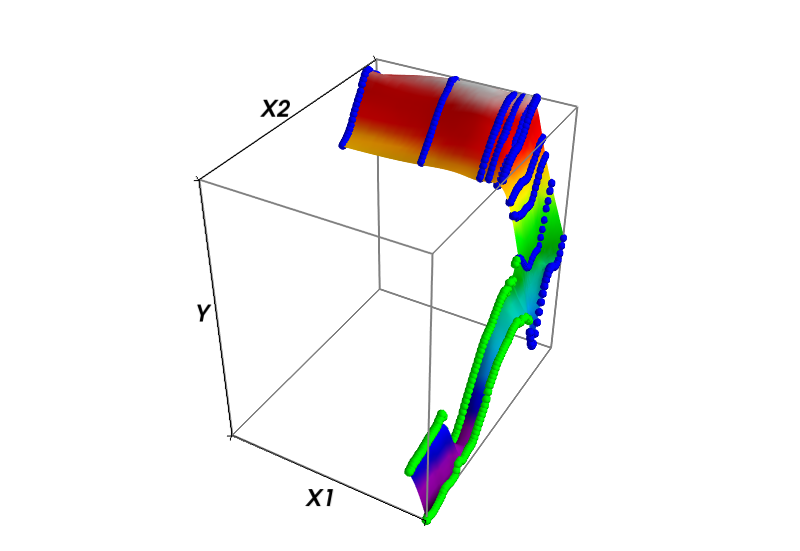}
      \caption{iTA}
    \end{subfigure}
    \begin{subfigure}[b]{0.48\textwidth}
      \includegraphics[width=1.0\textwidth, clip, trim=60mm 0mm 70mm 10mm]{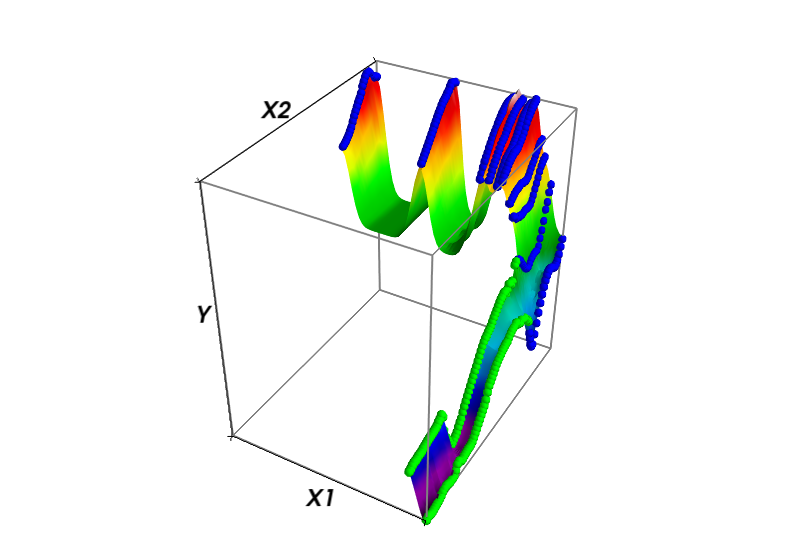}
      \caption{GP}
    \end{subfigure}
    \caption{iTA and GP approximations in the aerodynamic problem.}
    \label{fig:ita_sgp_slices}
\end{figure}

In Figure \ref{fig:ita_sgp_slices} we compare, on a 2D slice of the region of main interest, the iTA surrogate model with a model obtained using the scikit-learn Gaussian Process technique \cite{Sklearn2011}, which may be considered as a conventional approach for this problem (since the full DoE is not factorizable). We observe physically unnatural ``valleys'' in the GP model. This degeneracy results from the GP's assumptions of uniformity and homogeneity of data \cite{Rasmussen2005} that do not hold in this problem due to gaps in the DoE and large gradient of the response function in a part of the design space.  Clearly, the iTA model does not have these drawbacks. In addition, iTA is much faster to train on this $2026$-point set: it took 10 seconds for the iTA model and $1800$ seconds for the GP model\footnote{The experiments were conducted on a PC with Intel(R) Core(TM) i7-2600 CPU @ 3.40GHz and 8GB RAM.}.

\section{Conclusion}

We have described GTApprox --- a new tool for medium-scale surrogate modeling in industrial design -- and its novel features that make it convenient for surrogate modeling, especially for applications in engineering and for use by non-experts in data analysis. The tool contains some entirely new approximation algorithms (e.g., Tensor Approximation with arbitrary factors and incomplete Tensor Approximation) as well as novel model selection meta-algorithms. In addition, GTApprox supports multiple novel ``non-technical'' options and features allowing the user to more easily express the desired properties of the model or some domain-specific properties of a data. 

When compared to scikit-learn algorithms in the default mode on a collection of test problems, GTApprox shows a superior accuracy. This is achieved at the cost of longer training times that, nevertheless, remain moderate for medium-scale problems.    

We have also briefly described a few applications of GTApprox to real engineering problems where a crucial role was played by the tool's distinctive elements (the new algorithms MoA and iTA, automated model selection, built-in availability of gradients).

\section*{Acknowledgments}
The scientific results of Sections \ref{sec:algo} and \ref{sec:hyper_sel} are based on the research conducted at IITP RAS and supported by the Russian Science Foundation (project 14-50-00150).

\section*{References}

\bibliographystyle{model1-num-names}
\bibliography{gtapprox_refs}

\end{document}